\documentclass[conference]{IEEEtran}
\IEEEoverridecommandlockouts

\usepackage{cite}
\usepackage{amsmath,amssymb,amsfonts}
\usepackage{algorithmic}
\usepackage{graphicx}
\usepackage{textcomp}
\usepackage{xcolor}
\usepackage{booktabs}
\usepackage{multirow}
\usepackage{url}
\usepackage{tikz}
\usepackage{pgfplots}
\pgfplotsset{compat=1.17}
\usetikzlibrary{shapes.geometric, arrows.meta, positioning, fit, backgrounds, calc}

\def\BibTeX{{\rm B\kern-.05em{\sc i\kern-.025em b}\kern-.08em
    T\kern-.1667em\lower.7ex\hbox{E}\kern-.125emX}}

\tikzset{
  box/.style={
    rectangle, draw, rounded corners=2pt,
    minimum width=3cm, minimum height=0.7cm,
    text centered, align=center,
    font=\small
  },
  gbox/.style={
    rectangle, draw, rounded corners=2pt,
    minimum width=3cm, minimum height=0.7cm,
    text centered, align=center,
    font=\small, fill=gray!15
  },
  wbox/.style={
    rectangle, draw, rounded corners=2pt,
    minimum width=3cm, minimum height=0.7cm,
    text centered, align=center,
    font=\small, fill=orange!10
  },
  diam/.style={
    diamond, draw,
    minimum width=2cm, minimum height=0.8cm,
    text centered, align=center,
    font=\small, aspect=2
  },
  rdiam/.style={
    diamond, draw,
    minimum width=1.5cm, minimum height=0.7cm,
    text centered, align=center,
    font=\small, aspect=2, fill=yellow!20
  },
  label/.style={
    font=\small\itshape, align=center
  },
  arr/.style={
    ->, >=Stealth, thick
  },
  succarr/.style={
    ->, >=Stealth, thick, green!60!black
  }
}

\begin{document}

\title{Multi-Agent Retrieval-Augmented Generation\\
for Efficient Cloud Knowledge Base Search in Telecom SNOC Environment}

\author{
\IEEEauthorblockN{Harish Saragadam}
\IEEEauthorblockA{\textit{Vodafone Idea -- SNOC}\\
harish.saragadam@vodafoneidea.com}
\and
\IEEEauthorblockN{Sudhanshu Sharma}
\IEEEauthorblockA{\textit{Vodafone Idea -- SNOC}\\
sudhanshu.sharma@vodafoneidea.com}
\and
\IEEEauthorblockN{Ipsha Routray}
\IEEEauthorblockA{\textit{Vodafone Idea -- SNOC}\\
ipsha.routray1@vodafoneidea.com}
}

\maketitle

\begin{abstract}
 Telecom Service and Network Operations Centers (SNOCs) depend on large collections of cloud documents, such as Standard Operating Procedures (SOPs), vendor technical bulletins, incident post-mortem reports, and network configuration guides, to ensure uninterrupted network services. During network incidents, quickly finding accurate information is essential. However, traditional keyword-based search and single-stage retrieval methods often do not provide the level of accuracy needed in such time-critical situations.  
 
 This paper introduces Athena for cloud Knowledge Base, a fully offline, multi-agent Retrieval-Augmented Generation (RAG) pipeline developed specifically for cloud document search at Vodafone Idea's SNOC. The system combines dense vector retrieval using E5-Large-V2 (1024-dimensional embeddings), BM25 sparse keyword search, and Knowledge Graph (KG) expansion within a LangGraph multi-agent orchestration framework. The ranked results from these three retrieval methods are combined using Weighted Combination Sum (CombSUM). A cross-encoder reranker and Maximal Marginal Relevance (MMR) are then used to select a diverse and highly relevant set of evidence.

To improve the reliability of the generated responses, the system performs a per-chunk LLM evaluation with an explicit attribution verifier, evaluating each MMR-selected chunk one at a time to remove unsupported or irrelevant answers before generating the final response. If none of the individual chunks produces a valid answer, the system automatically falls back to evaluating multiple chunks together.
The proposed system was evaluated on a corpus of 4,200 SNOC cloud documents (approximately 312,000 indexed chunks). DocuSearch achieved a Mean Reciprocal Rank (MRR@10) of 0.910 and an Exact Match (EM) accuracy of 78.4\%,
representing a 14.6-point improvement over single-stage dense retrieval. The entire pipeline operates on-premises without any inference-time internet calls, meeting telecom-grade data sovereignty requirements.

\end{abstract}


%=============================================================================
\section{Introduction}
%=============================================================================

\subsection{Background}
Modern telecom IP/MPLS networks generate vast operational knowledge stored across thousands of cloud documents: vendor datasheets, SOP manuals, infrastructure runbooks, and historical incident tickets. Engineers at SNOCs must rapidly locate authoritative information during active outages, often consulting multiple document types simultaneously. The shift toward multi-vendor, cloud-native, and 5G architectures has dramatically increased both document volume and query complexity.

\subsection{Problem Statement}
Existing information retrieval tools deployed at telecom SNOCs are inadequate for three reasons. First, keyword-search engines miss semantically equivalent phrasings and cannot rank by contextual relevance. Second, single-stage dense retrieval systems miss lexically exact matches critical in telecom contexts--router model numbers, alarm codes, interface identifiers, and vendor-specific error strings. Third, LLM-based generation without explicit attribution verification introduces unsupported claims that can mislead engineers during time-critical incidents.

\subsection{Research Objectives and Contributions}
This work designs a three-source hybrid retrieval system (dense +
BM25 + KG) with CombSUM fusion, builds a LangGraph multi-agent
orchestration framework, and implements a per-chunk attribution
verification loop---all validated on a real-world SNOC corpus.
Specific contributions are:
\begin{itemize}
  \item Three-source retrieval with weighted CombSUM
        ($\omega_v{=}0.50$, $\omega_b{=}0.35$, $\omega_k{=}0.15$,
        $\sum_s\omega_s{=}1$), where $v$, $b$, $k$ index vector,
        BM25, and KG sources respectively.
  \item Cascaded domain-aware agent (regex gazetteer + LLM) for telecom network, cloud, hardware, and operational document repositories.
  \item Per-chunk loop: neighbor-need detection $\to$ sufficiency
        check $\to$ generation $\to$ attribution verification,
        with multi-chunk fallback.
  \item Five answer templates with template-specific prompting:
        \texttt{procedure\_sop}, \texttt{concept\_explanation},
        \texttt{comparison\_decision}, \texttt{parameter\_lookup},
        \texttt{troubleshooting\_rca}.
  \item Ablation study on 4{,}200 SNOC documents isolating each
        component's contribution to end-to-end accuracy.
\end{itemize}

%=============================================================================
\section{Related Work}
%=============================================================================

\textbf{Traditional IR.} BM25~\cite{robertson1976}is one of the most widely used search methods due to its strong performance on keyword-based queries and solid theoretical foundation. However, it relies on exact word matching, making it less effective for semantic or paraphrased queries. It also retrieves relevant documents but does not generate concise, synthesized answers.

\textbf{Dense Retrieval.} Bi-encoder dense retrieval models~\cite{karpukhin2020} enable sub-millisecond ANN search. E5-Large-V2~\cite{wang2022} achieves state-of-the-art results on MTEB by applying text-to-text contrastive training with weakly supervised data, producing 1,024-dimensional embeddings that excel at both symmetric and asymmetric retrieval. Dense models capture paraphrase-heavy queries effectively but can underperform BM25 on exact-match patterns prevalent in telecom documents.

\textbf{Score Fusion.} CombSUM~\cite{fox1993} aggregates normalised document scores across multiple retrieval sources by simple weighted summation. Unlike rank-based fusion methods, CombSUM exploits the magnitude of individual retrieval scores, rewarding documents that achieve high absolute scores across multiple sources rather than merely high ranks. This property is particularly beneficial in hybrid retrieval settings where lexical and semantic relevance signals carry complementary information.

\textbf{RAG.} Lewis et al.~\cite{lewis2020}  introduced Retrieval-Augmented Generation (RAG), where LLMs generate responses using retrieved evidence to improve factual grounding. Fusion-in-Decoder (FiD) [6] extends this idea by jointly attending to multiple retrieved passages, while Self-RAG [7] adds adaptive retrieval and self-reflection to improve response quality. Although effective, these methods do not consider hybrid retrieval from multiple sources or iterative chunk-level attribution verification.

\textbf{Multi-Agent Systems.} LangGraph~\cite{langgraph2024} models LLM applications as directed state graphs with conditional execution, enabling reliable multi-agent workflows. Most existing work focuses on open-domain QA, whereas our approach targets offline enterprise document search with strict latency and data-sovereignty constraint.

\textbf{Research Gap.} No prior work addresses: (a) three-source hybrid retrieval for telecom cloud documents using CombSUM fusion; (b) cascaded vendor detection with regex gazetteer and LLM confirmation; (c) per-chunk attribution verification with neighbor expansion and explicit fallback; and (d) fully offline deployment with five answer-type templates.

%=============================================================================
\section{Dataset Preparation}
%=============================================================================

\subsection{Data Sources}
The SNOC cloud document corpus is compiled from six heterogeneous sources: internal SOP repositories, network infrastructure documentation, cloud platform administration guides, operational runbooks, telecommunications equipment manuals, and historical incident post-mortem reports extracted from the SNOC ticketing system.

\subsection{Document Collection and Cleaning}
A total of 4,200 PDF documents were collected (approximately 68,000 pages). Categories span: SOPs (31\%), vendor technical bulletins (24\%), incident post-mortems (19\%), configuration guides (14\%), regulatory compliance manuals (7\%), and topology/capacity planning documents (5\%). Each document is processed by Docling for layout detection, section identification, table extraction, and image isolation. For scanned or image-heavy pages, RapidOCR provides fully local OCR fallback.

\subsection{Chunking and Metadata}
Hierarchical, section-aware chunking produces semantically coherent chunks with target length 512 tokens and 10\% overlap. Each chunk inherits its parent section heading, document ID, and page number as metadata. The corpus yields 312,000 chunks. Critically, each chunk records \texttt{prev\_chunk\_id} and \texttt{next\_chunk\_id} pointers, enabling the per-chunk evaluator to fetch neighboring context during evaluation. Embeddings are computed using E5-Large-V2 (1,024-dimensional, GPU-accelerated via CUDA, batch size 64) and stored in a Qdrant HNSW index ($ef_\text{construction}=200$, $M=16$). Importantly, E5-Large-V2 requires query inputs to be prefixed with \texttt{"query: "} and passage inputs with \texttt{"passage: "} at inference time, following the model's asymmetric text-to-text training protocol. A BM25Okapi index is built over the tokenized corpus and serialized as \texttt{bm25\_index.pkl}.The complete indexing process takes approximately 4.2 hours.
%=============================================================================
\section{System Architecture}
%=============================================================================

\subsection{Overview}
DocuSearch consists of two phases: an offline ingestion pipeline and a real-time query pipeline orchestrated as a LangGraph \texttt{StateGraph}. The query pipeline is split into two functional phases:

\begin{itemize}
    \item \textbf{Pre-MMR} (\texttt{run\_supervisor}): Query Understanding, BM25 Keyword Extraction, Vendor Detection, three parallel retrieval streams, CombSUM Fusion, Cross-Encoder Reranking, and MMR Evidence Selection.
    \item \textbf{Post-MMR} (\texttt{run\_per\_chunk\_evaluation}): Per-chunk neighbor-need detection, sufficiency checking, answer generation, and attribution verification, with a multi-chunk fallback path.
\end{itemize}

\subsection{User Interface Layer}
The UI is a Dash-based Python web application running on an internal server. It exposes seven tabs: \textit{Answer} (LLM response with inline citations), \textit{Evidence} (retrieved chunk viewer with source metadata), \textit{Tables} (extracted structured tables), \textit{Images} (document page thumbnails with image-quality filtering), \textit{Timeline} (per-agent latency trace), \textit{Trace} (full per-agent state log), and \textit{Follow-ups} (auto-generated follow-up questions). An ad-hoc upload mode allows engineers to query arbitrary PDF documents without full corpus indexing.

\subsection{Query Understanding Agent (Template Selector)}
Each incoming query is classified into one of five \textit{answer templates}: \texttt{procedure\_sop} (step-by-step tasks), \texttt{concept\_explanation} (definitions and overviews), \texttt{comparison\_decision} (side-by-side comparisons), \texttt{parameter\_lookup} (specific values, specs, or table data), and \texttt{troubleshooting\_rca} (fault diagnosis and corrective action). Classification uses a lightweight LLM call; a heuristic regex fallback activates if the LLM is unavailable or returns an unrecognized label. The selected template governs both BM25 keyword extraction strategy and the generation prompt format.

\subsection{BM25 Keyword Extraction Agent}
Concurrent with template classification, a second LLM call extracts high-signal query terms for BM25 sparse retrieval. Results are filtered against a domain-keyword vocabulary of telecom terms  and a stop-word list. If LLM extraction fails, a deterministic telecom-domain fallback regex extracts candidate terms directly.

\subsection{Vendor Detection Agent}
The Vendor Detection Agent (VDA) identifies vendor entities in the query using a three-stage cascaded strategy:
\begin{enumerate}
    \item \textbf{Regex gazetteer sweep}: A fast O(\textit{vendors} $\times$ \textit{aliases}) pass over all vendors using both vendor folder names and their underscore-split parts. Always runs.
    \item \textbf{LLM-guided detection}: If vLLM is active, a structured JSON-output prompt classifies the query against the vendor list with inline abbreviation annotations. Otherwise, a local indexed LLM call returns vendor indices.
    \item \textbf{Merge}: LLM and regex results are unioned; detected vendors scope Qdrant ANN search via metadata filters, pre-filter the BM25 index to vendor-tagged chunks, and anchor KG expansion to vendor-specific entity nodes. If no vendor is detected, exhaustive search across all vendors is performed.
\end{enumerate}
The vendor registry is built at startup from the \texttt{documents} SQLite table by extracting the first directory component of each document's source path.

\subsection{Retrieval Agent}
Three parallel retrieval sub-tasks execute concurrently:
\begin{itemize}
    \item \textbf{Vector search}: Qdrant ANN with HNSW ($ef=128$), top-40, with optional vendor-metadata filter. Query vectors are produced with the \texttt{"query: "} prefix required by E5-Large-V2. Latency $\approx 8$~ms.
    \item \textbf{BM25 retrieval}: BM25Okapi keyword search over SQLite FTS5 index, top-40, vendor-scoped when applicable. Latency $\approx 12$~ms.
    \item \textbf{KG expansion}: SQLite FTS5 traversal from seed chunks (top-20 vector + top-20 BM25), top-15 neighbor chunks. Latency $\approx 5$~ms.
\end{itemize}

\begin{figure}
\centering
\begin{tikzpicture}[
  node distance=0.45cm and 0.8cm,
  box/.style={draw=blue!80!black, fill=blue!10, rounded corners=2pt, minimum width=3.4cm, minimum height=0.55cm, font=\scriptsize\sffamily, align=center, thick},
  wbox/.style={draw=orange!80!black, fill=orange!15, rounded corners=2pt, minimum width=3.4cm, minimum height=0.55cm, font=\scriptsize\sffamily, align=center, thick},
  gbox/.style={draw=green!60!black, fill=green!15, rounded corners=2pt, minimum width=3.4cm, minimum height=0.55cm, font=\scriptsize\sffamily, align=center, thick},
  diam/.style={draw=yellow!80!black, fill=yellow!10, diamond, aspect=2.2, minimum width=2.5cm, minimum height=0.65cm, font=\scriptsize\sffamily, align=center, inner sep=0pt, thick},
  rdiam/.style={draw=red!70!black, fill=red!5, diamond, aspect=2.2, minimum width=1.8cm, minimum height=0.55cm, font=\tiny\sffamily, align=center, inner sep=0pt, thick},
  arr/.style={-{Stealth[length=4.5pt]}, thick, draw=black!90},
  succarr/.style={-{Stealth[length=4.5pt]}, thick, draw=green!60!black},
  dasharr/.style={-{Stealth[length=4.5pt]}, dashed, red!80!black, thick},
  dotarr/.style={-{Stealth[length=4.5pt]}, dotted, orange!95!black, ultra thick},
  label/.style={font=\tiny\sffamily\bfseries, text=black!80}
]

%% --- PRE-MMR PHASE ---
\node[box] (step1) {STEP 1:\\Template Selection};
\node[box, below=of step1] (step2) {STEP 2:\\BM25 Keyword Extraction};
\node[box, below=of step2] (step2b) {STEP 2b:\\Vendor Detection};

\node[box, minimum width=1.8cm, below left=0.6cm and -1.0cm of step2b] (step3) {STEP 3\\Vector Search};
\node[box, minimum width=1.8cm, right=0.2cm of step3] (step4) {STEP 4\\BM25 Retrieval};
\node[box, minimum width=1.8cm, right=0.2cm of step4] (step5) {STEP 5\\KG Expansion};

\node[box, below=0.6cm of step4] (step5b) {STEP 5b:\\CombSUM Fusion};
\node[box, below=of step5b] (step6) {STEP 6:\\Cross-Encoder Reranking};
\node[box, below=of step6] (step7) {STEP 7:\\MMR Selection};

\draw[arr] (step1) -- (step2);
\draw[arr] (step2) -- (step2b);
\draw[arr] (step2b.south) -- ++(0,-0.25) -| (step3.north);
\draw[arr] (step2b.south) -- ++(0,-0.25) -| (step4.north);
\draw[arr] (step2b.south) -- ++(0,-0.25) -| (step5.north);
\draw[arr] (step3.south) |- (step5b.west);
\draw[arr] (step4.south) -- (step5b.north);
\draw[arr] (step5.south) |- (step5b.east);
\draw[arr] (step5b) -- (step6);
\draw[arr] (step6) -- (step7);

%% Divider Line
\node[label, below=0.5cm of step7] (divlabel) {\textit{ordered MMR chunks $\downarrow$}};
\draw[arr] (step7) -- (divlabel);

%% --- POST-MMR PHASE ---
\node[box, below=0.5cm of divlabel] (pstep2) {STEP 2 (per-chunk):\\LLM - Current Chunk};
\node[diam, below=0.5cm of pstep2] (needmore) {Need more\\context?};
\node[gbox, right=0.5cm of needmore, minimum width=3.1cm] (fetchnbr) {Fetch Neighboring\\Chunk Nearby};

\node[box, below=0.5cm of needmore] (pstep3) {STEP 3:\\LLM + Chunk (current only)};
\node[diam, below=0.5cm of pstep3] (sufficient) {Sufficient?};
\node[box, below=0.5cm of sufficient] (pstep4) {STEP 4:\\LLM Answer Generation};
\node[box, below=0.45cm of pstep4] (pstep5) {STEP 5:\\    Attribution Verifier LLM};
\node[diam, below=0.5cm of pstep5] (correct) {Answer\\correct?};
\node[gbox, fill=green!15, below=0.5cm of correct] (terminate) {\textbf{TERMINATE}\\method = per\_chunk};

%% Outer Iteration Decision Diamonds
\node[rdiam, left=1.2cm of sufficient] (moreleft) {More\\chunks?};
\node[rdiam, right=1.2cm of correct] (moreright) {More\\chunks?};

%% Fallback Layer Nodes
\node[wbox, below=0.9cm of terminate] (fallback) {FALLBACK:\\Merge All Expanded Contexts};
\node[wbox, below=0.4cm of fallback] (fbstep4) {Fallback STEP 4:\\LLM Multi-chunk Answer};
\node[wbox, below=0.4cm of fbstep4] (fbstep5) {Fallback STEP 5:\\Attribution Check};
\node[wbox, below=0.4cm of fbstep5] (retans) {\textbf{Return Answer}\\method = fallback\_merge};

%% Center Stream Sequential Connections
\draw[arr] (divlabel) -- (pstep2);
\draw[arr] (pstep2) -- (needmore);
\draw[arr] (needmore) -- node[label,right,pos=0.3] {No} (pstep3);
\draw[succarr] (needmore) -- node[label,above,pos=0.4] {Yes} (fetchnbr);
\draw[arr] (pstep3) -- (sufficient);
\draw[succarr] (sufficient) -- node[label,right,pos=0.3] {Yes} (pstep4);
\draw[arr] (pstep4) -- (pstep5);
\draw[arr] (pstep5) -- (correct);
\draw[succarr] (correct) -- node[label,right,pos=0.3] {Yes} (terminate);
\draw[arr] (fallback) -- (fbstep4);
\draw[arr] (fbstep4) -- (fbstep5);
\draw[arr] (fbstep5) -- (retans);

%% --- HIGH-PRECISION NON-OVERLAPPING ROUTING ---

%% 1. Upper Success Loop: Directly routes up and enters STEP 2 from the right face
\draw[succarr] (fetchnbr.north) |- (pstep2.east);

%% 2. Left Outer Feedback Path
\draw[dasharr] (sufficient.west) -- node[label,above,pos=0.5] {No} (moreleft.east);
\draw[dasharr] (moreleft.north) |- node[label,left,pos=0.25] {Yes} (pstep2.west);
\draw[dotarr] (moreleft.south) |- node[label,left,pos=0.2] {No} (fallback.west);

%% 3. Right Outer Feedback Path: Sweeps wide around everything and feeds into the top arrow entry point
\draw[dasharr] (correct.east) -- node[label,above,pos=0.5] {No} (moreright.west);
\draw[dasharr] (moreright.east) -- ++(1.6,0) node[label,above,pos=0.15] {Yes} |- ([yshift=0.25cm]pstep2.north) -- (pstep2.north);
\draw[dotarr] (moreright.south) |- node[label,right,pos=0.2] {No} (fallback.east);

\end{tikzpicture}
\caption{DocuSearch multi-agent query pipeline diagram matching architectural specification layout.}
\label{fig:pipeline}
\end{figure}
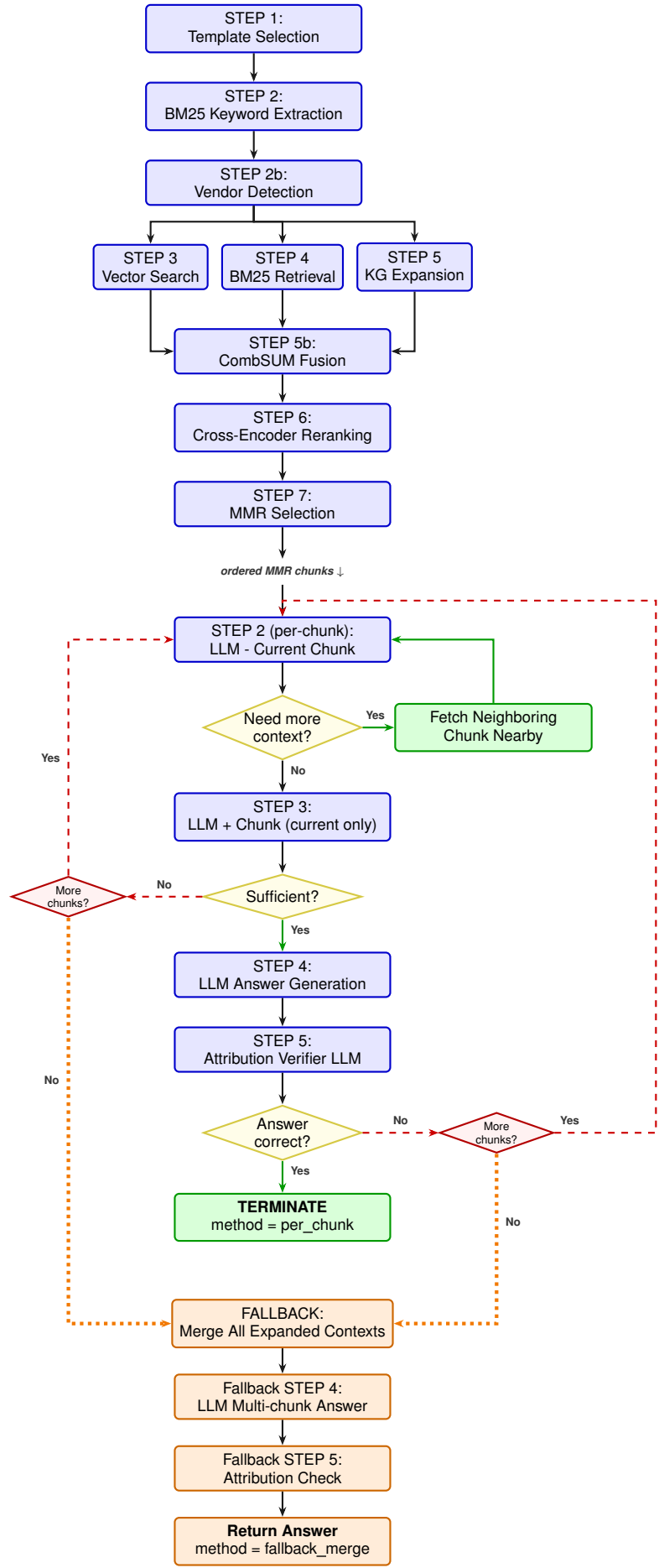

The complete pipeline architecture is illustrated in Fig.~\ref{fig:pipeline}.

\subsection{Ranking Agent}
Weighted CombSUM merges the three scored lists by normalising each source's scores to $[0,1]$ via min-max scaling, then computing a weighted sum: $\omega_v=0.50$, $\omega_b=0.35$, $\omega_k=0.15$, $\sum\omega=1$. Unlike rank-based fusion, CombSUM leverages absolute score magnitudes and therefore benefits from the calibrated cosine similarities produced by E5-Large-V2 and the well-normalised BM25 scores. The fused list is sorted descending by CombSUM score to yield a unified top-$N$ candidate list. A cross-encoder reranker then jointly scores each query--candidate pair using a 6-layer sentence-transformers model running on CUDA, reducing to top-12. MMR selection with $\lambda=0.65$ yields the final ordered 10 evidence chunks. Cross-encoder reranking dominates pre-MMR latency at $\approx 380$~ms.

\subsection{Per-Chunk Evaluation Agent}
This agent replaces the naive joint-context generation approach with an independent evaluation loop over each MMR chunk:

\textbf{Step 1 -- Neighbor-need detection}: An LLM prompt assesses whether the anchor chunk provides sufficient context or requires neighboring chunks. If context is needed, \texttt{prev\_chunk\_id} and \texttt{next\_chunk\_id} pointers are used to fetch neighbors (up to \texttt{MAX\_NEIGHBOR\_LOOPS=2} iterations). The LLM may also flag a chunk as off-topic (\texttt{skip}), advancing immediately to the next MMR chunk.

\textbf{Step 2 -- Sufficiency check}: The expanded chunk context (anchor + any fetched neighbors) is evaluated by an LLM for sufficiency to answer the query. Insufficient chunks are skipped.

\textbf{Step 3 -- Answer generation}: Mistral-7B-Instruct-v0.2 via vLLM generates a template-specific answer from the expanded context only, with citation markers referencing source chunk and document. Temperature is set to 0.1 for grounded, low-variance outputs.

\textbf{Step 4 -- Attribution verification}: A chain-of-thought LLM prompt breaks the generated answer into simple sentences and checks whether each one can be traced back to a specific span in the source evidence, assigning a binary attribution label to each sentence. If the overall attribution score is below 0.70, the answer is rejected and the system moves to the next MMR chunk. Unlike NLI-based groundedness methods, this approach requires explicit linking to evidence spans, making verification more precise and easier for SNOC engineers to interpret.

\textbf{Fallback path}: If all 10 MMR chunks are exhausted without a fully attributed answer, all per-chunk expanded contexts are deduplicated and merged. A final LLM generation and attribution check run over the merged context (\texttt{method=fallback\_merge}), producing the final answer regardless of attribution score.

\subsection{Response Generation Agent}
Five template-specific prompts govern generation format: \texttt{procedure\_sop} produces numbered steps; \texttt{parameter\_lookup} produces value/table/notes structure; \texttt{comparison\_decision} produces side-by-side structure; \texttt{concept\_explanation} produces definition/overview/application structure; \texttt{troubleshooting\_rca} produces fault analysis and remediation steps. All templates require inline citation markers referencing source chunks. The answer is cleaned of internal chain-of-thought artifacts before delivery.

%=============================================================================
\section{Mathematical Formulation}
%=============================================================================

Throughout this section: $c_i$ denotes the $i$-th indexed chunk;
$Q$ the user query; $\phi_q(\cdot)$ and $\phi_p(\cdot)$ the
E5-Large-V2 bi-encoder applied in query mode (prefix
\texttt{query:}) and passage mode (prefix \texttt{passage:}),
respectively; and $\|\cdot\|_2$ the Euclidean norm.

\subsection{E5-Large-V2 Embeddings}
E5-Large-V2 requires asymmetric prefix tokens to preserve the
directional relevance signal learned during contrastive
pre-training. Query and passage vectors are mapped to unit-norm
embeddings in $\mathbb{R}^{1024}$:
\begin{align}
  \mathbf{q}   &= \phi_q(Q),   \quad \mathbf{q}\in\mathbb{R}^{1024},\; \|\mathbf{q}\|_2=1,
  \label{eq:qemb}\\
  \mathbf{d}_i &= \phi_p(c_i), \quad \mathbf{d}_i\in\mathbb{R}^{1024},\; \|\mathbf{d}_i\|_2=1.
  \label{eq:demb}
\end{align}
Passage embeddings $\mathbf{d}_i$ are computed offline during
ingestion~(Eq.~\ref{eq:demb}); query embeddings $\mathbf{q}$ are
computed at query time~(Eq.~\ref{eq:qemb}). Under
$\ell_2$-normalisation, cosine similarity reduces to the dot
product:
\begin{equation}
  s_{\mathrm{vec}}(c_i,Q)=\mathbf{d}_i\cdot\mathbf{q}\;\in[-1,\,1].
  \label{eq:svec}
\end{equation}

\subsection{BM25 Scoring}
Let $t$ range over query terms; $f(t,c_i)$ the term frequency of
$t$ in chunk~$c_i$; $n(t)$ the number of chunks containing~$t$;
$N$ the total chunk count; $|c_i|$ the chunk length in tokens;
and $\mathrm{avgdl}$ the corpus-mean chunk length in tokens:
\begin{equation}
  s_{\mathrm{bm25}}(c_i,Q)=\sum_{t\in Q}\mathrm{IDF}(t)\cdot
  \frac{f(t,c_i)(k_1+1)}{f(t,c_i)+k_1\!\left(1-b+b\dfrac{|c_i|}{\mathrm{avgdl}}\right)},
  \label{eq:bm25}
\end{equation}
with $k_1{=}1.5$, $b{=}0.75$, and the Robertson--Sp\"{a}rck Jones
IDF variant used by BM25Okapi:
\begin{equation}
  \mathrm{IDF}(t)=\log\!\frac{N-n(t)+0.5}{n(t)+0.5}+1.
  \label{eq:idf}
\end{equation}
The additive constant $+1$ in~\eqref{eq:idf} ensures
non-negative IDF scores for all terms.

\subsection{Weighted CombSUM Fusion}
\label{sec:combsum}
Let $s\in\{v,b,k\}$ index the vector, BM25, and KG retrieval
sources, and let $R_s$ denote the candidate set returned by
source~$s$ ($|R_v|{=}|R_b|{=}40$; $|R_k|{=}15$). Each raw
score is min-max normalised over $R_s$:
\begin{equation}
  \hat{s}_s(c)=
  \frac{s_s(c)-\displaystyle\min_{c'\in R_s}s_s(c')}
       {\displaystyle\max_{c'\in R_s}s_s(c')-\min_{c'\in R_s}s_s(c')+\varepsilon},
  \quad\varepsilon=10^{-9}.
  \label{eq:norm}
\end{equation}
Chunks absent from $R_s$ are assigned $\hat{s}_s(c)=0$. The
weighted CombSUM score aggregates the three normalised signals:
\begin{equation}
  \mathrm{CombSUM}(c)=\omega_v\,\hat{s}_v(c)
                     +\omega_b\,\hat{s}_b(c)
                     +\omega_k\,\hat{s}_k(c),
  \label{eq:combsum}
\end{equation}
with $\omega_v{=}0.50$, $\omega_b{=}0.35$, $\omega_k{=}0.15$,
and $\sum_s\omega_s{=}1$. Assigning zero to absent-source chunks
preserves an additive reward for candidates retrieved by multiple
sources.

\subsection{MMR Evidence Selection}
\label{sec:mmr}
Let $C'$ be the top-12 cross-encoder-reranked candidates, $S$ the
set of already-selected evidence chunks (initially $S=\emptyset$),
and $s_{\mathrm{CE}}(c,Q)$ the cross-encoder relevance score for
chunk~$c$ given query~$Q$. MMR iteratively appends
\begin{equation}
  c^*=\!\arg\max_{c\in C'\setminus S}\!
  \Bigl[\lambda\cdot s_{\mathrm{CE}}(c,Q)
        -(1{-}\lambda)\!\max_{c_j\in S}\mathbf{d}_c\!\cdot\!\mathbf{d}_{c_j}\Bigr],
  \label{eq:mmr}
\end{equation}
until $|S|{=}10$, with $\lambda{=}0.65$. When $S{=}\emptyset$ the
diversity term is defined as zero (i.e.~the first chunk is always
selected by relevance alone). The inner product
$\mathbf{d}_c\!\cdot\!\mathbf{d}_{c_j}$ equals the cosine
similarity between candidate~$c$ and already-selected chunk~$c_j$
because all E5-Large-V2 passage embeddings are unit-norm
(Eq.~\ref{eq:demb}).

\subsection{Attribution Scoring}
\label{sec:attr}
Let $A$ be the generated answer, $S_A{=}\{\sigma_1,\sigma_2,\ldots\}$
its decomposition into atomic sentences, and $E$ the expanded
evidence for the current MMR chunk. The attribution score is:
\begin{equation}
  \mathcal{A}(A,E)=\frac{1}{|S_A|}
  \sum_{\sigma\in S_A}
  \mathbf{1}\!\left[\,\exists\,e\in E:\mathrm{Attr}(\sigma,e)=1\,\right],
  \label{eq:attr}
\end{equation}
where $\mathrm{Attr}(\sigma,e){=}1$ if the LLM identifies a
supporting evidence span in~$e$ for sentence~$\sigma$, and $0$
otherwise. Here $\mathcal{A}$ is the scoring function (distinct
from the answer~$A$). Answers with $\mathcal{A}(A,E)<\tau{=}0.70$
are rejected. Unlike soft NLI entailment,
Eq.~\eqref{eq:attr} requires the model to produce a verifiable
evidence pointer, yielding directly auditable answer--source links.
%=============================================================================
\section{Experimental Results}
%=============================================================================

\subsection{Experimental Setup}
All experiments were performed in a local offline environment with consistent computational resources across all evaluations. Latency measurements are averaged over 250 held-out test queries (5 runs, mean ± standard deviation).

\subsection{Evaluation Metrics}
MRR@10 (Mean Reciprocal Rank), NDCG@10, EM (Exact Match against gold annotation), F1 (token-overlap), Precision@5, Recall@5, and end-to-end wall-clock latency.

\subsection{Comparative Study}
\begin{table}[h]
\centering
\caption{ SNOC Document QA Benchmark ($n{=}250$ queries).
         All vector-based baselines use E5-Large-V2.
         EM and F1 in \%.}
\label{tab:main}
\begin{tabular}{@{}lccccc@{}}
\toprule
\textbf{System} & \textbf{MRR@10} & \textbf{NDCG@10} & \textbf{EM} & \textbf{F1} & \textbf{Lat.(s)} \\
\midrule
Dense-Only       & 0.761 & 0.703 & 63.8 & 71.2 & 0.42 \\
BM25-Only        & 0.712 & 0.668 & 59.4 & 67.8 & 0.18 \\
CombSUM (no rerank) & 0.812 & 0.779 & 68.2 & 75.4 & 0.55 \\
CombSUM + CE     & 0.871 & 0.843 & 73.6 & 80.1 & 1.20 \\
\textbf{DocuSearch} & \textbf{0.910} & \textbf{0.887} & \textbf{78.4} & \textbf{84.7} & 2.15 \\
\bottomrule
\end{tabular}
\end{table}

\subsection{Ablation Analysis}\begin{table}[h]
\centering
\caption{ Ablation results. $\Delta$EM in percentage points
         relative to Full DocuSearch; rows ordered by impact }
\label{tab:ablation}
\begin{tabular}{@{}lcc@{}}
\toprule
\textbf{Configuration} & \textbf{EM\%} & \textbf{$\Delta$EM} \\
\midrule
Full DocuSearch               & 78.4 & --- \\
w/o Attribution Verification  & 72.1 & $-6.3$ \\
w/o Per-Chunk Evaluation Loop & 71.8 & $-6.6$ \\
w/o KG Expansion              & 74.9 & $-3.5$ \\
w/o Cross-Encoder Reranking   & 73.6 & $-4.8$ \\
w/o MMR (top-10 by CE score)  & 76.2 & $-2.2$ \\
w/o Vendor Detection          & 75.8 & $-2.6$ \\
\bottomrule
\end{tabular}
\end{table}

\subsection{Latency Decomposition}
End-to-end latency is 2.15~s. Stage-wise breakdown: ANN vector search ($\approx$8~ms), BM25 retrieval ($\approx$12~ms), KG expansion ($\approx$5~ms), CombSUM fusion ($\approx$2~ms), cross-encoder reranking ($\approx$380~ms), MMR selection ($\approx$15~ms), LLM generation ($\approx$1.3~s). The cross-encoder and LLM together account for 78\% of total latency.

%=============================================================================
\section{Telecom Use Cases}
%=============================================================================

\textbf{Incident Resolution.} During a BGP neighbor-down event, an engineer queries: \textit{``BGP session reset SOP for a core router.''} DocuSearch's KG expansion retrieves the relevant SOP along with related configuration guides and historical incident records, helping reduce MTTR.

\textbf{Cloud Deployment.} Deployment checklist queries trigger the \texttt{procedure\_sop} template, generating step-by-step procedures with inline document references.

\textbf{Cloud Administration.} Cloud platform management and version upgrade queries are answered using the relevant administration guides. Comparative version queries invoke the \texttt{comparison\_decision} template.

\textbf{Network Troubleshooting.} Fault-related queries leverage KG expansion to retrieve troubleshooting documents and similar historical incidents, even when only a generic symptom is provided.

\textbf{Hardware Maintenance.} Queries related to firmware updates, server maintenance, and component replacement are matched to the appropriate hardware manuals using BM25 with exact model matching.

%=============================================================================
\section{Discussion}
%=============================================================================

\subsection{Result Analysis}
The ablation (Table~\ref{tab:ablation}) reveals that the per-chunk evaluation loop and attribution verifier are the most impactful components, contributing a combined $+12.9$ EM points over CombSUM+CE alone. This confirms that explicit span-level attribution checking is critical for safety-relevant SNOC workflows where engineers act on retrieved information under incident-response SLAs. KG expansion ($+3.5$ EM) is most impactful on entity-dense queries; vendor detection ($+2.6$ EM) shows its value on vendor-specific queries, which constitute 47\% of the test set.

\subsection{Key Design Novelties}
The cascaded vendor detection model combines a regex gazetteer sweep (always runs, zero LLM cost) with an LLM confirmation pass, handling both common cases efficiently and resolving ambiguous multi-vendor queries. The chunk-level neighbor-pointer architecture (prev/next IDs stored at ingestion time) enables the per-chunk evaluator to fetch coherent context without re-querying the vector index. The five-template answer system ensures that procedural queries produce actionable step-by-step outputs rather than prose summaries.

The choice of CombSUM over rank-based fusion is motivated by E5-Large-V2's well-calibrated cosine similarities: because the embedding model is trained with contrastive objectives that normalise output magnitudes, cosine scores carry ordinal and interval information that CombSUM can exploit---an advantage that purely rank-based methods discard. The asymmetric prefix protocol (\texttt{"query: "}/\texttt{"passage: "}) is enforced consistently at both indexing and query time to preserve the model's training distribution.

The span-level attribution verifier improves on NLI-based groundedness in two respects: it requires explicit evidence span identification rather than probabilistic entailment classification, and its binary per-sentence output is directly interpretable by engineers without calibration assumptions.

\subsection{Limitations}
The 2.15~s end-to-end latency may be unsuitable for real-time alert pipelines requiring sub-second responses. The cross-encoder and LLM together account for 78\% of latency and represent the primary optimization target. The KG is constructed with rule-based extraction that may miss implicit relations in informal incident post-mortems. CombSUM's performance is sensitive to score distribution mismatches between retrieval sources; outlier scores in any source can dominate the fused ranking if min-max normalisation is applied over small result sets.

%=============================================================================
\section{Future Work}
%=============================================================================

\textbf{Fine-tuned KG.} Replacing rule-based KG construction with a fine-tuned relation extraction model would improve recall on implicit relations, enabling richer multi-hop traversal for causal queries.

\textbf{Multimodal Retrieval.} SNOC documents contain network topology diagrams, interface tables, and KPI trend charts. A CLIP-based image encoder would retrieve relevant diagram pages alongside text chunks.

\textbf{Continuous Learning.} Engineer thumbs-up/thumbs-down feedback could fine-tune the template classifier and attribution verifier on a nightly basis.

\textbf{Telecom-Specific Embedding Fine-tuning.} Fine-tuning E5-Large-V2 on a SNOC-specific contrastive QA dataset would reduce embedding-space distance between synonymous telecom terms across vendors, further benefiting CombSUM fusion by producing better-calibrated per-source scores.

%=============================================================================
\section{Conclusion}
%=============================================================================

This paper presented Athena for cloud Knowledge Base, a fully offline multi-agent hybrid RAG system purpose-built for cloud document search in Vodafone Idea's SNOC environment. By combining dense vector retrieval (E5-Large-V2, 1,024-dimensional asymmetric embeddings), BM25 sparse search, and KG expansion under weighted CombSUM fusion ($\omega_v=0.50$, $\omega_b=0.35$, $\omega_k=0.15$), followed by cross-encoder reranking, MMR diversity selection ($\lambda=0.65$), and a per-chunk LLM evaluation loop with explicit attribution verification ($\tau=0.70$), DocuSearch achieves 78.4\% Exact Match accuracy---a 14.6-point improvement over single-stage dense retrieval---on a 312,000-chunk SNOC document corpus.

The ablation study demonstrates that the per-chunk evaluation loop and attribution verifier are the most impactful components ($+12.9$ EM combined), establishing that explicit span-level attribution checking is essential in safety-critical telecom operational workflows. The cascaded vendor-aware retrieval agent addresses the unique multi-vendor challenge of SNOC document corpora. The five-template answer generation system ensures operationally actionable outputs for each query type. The fully offline architecture satisfies data-sovereignty requirements under TRAI and ITU-T frameworks, making DocuSearch deployable at national telecom scale.

\end{document}